\documentclass[sigconf,natbib=true,anonymous=false]{acmart}

\usepackage{subfigure}
\usepackage{balance}

\usepackage{tabularx}

\acmYear{2024}
\acmConference[SIGIR'24]{ACM SIGIR Conference on Research and Development in Information Retrieval}{July 14--18, 2024}{Washington D.C., USA}

\copyrightyear{2024}
\acmYear{2024}
\setcopyright{acmlicensed}\acmConference[SIGIR '24]{Proceedings of the 47th
International ACM SIGIR Conference on Research and Development in
Information Retrieval}{July 14--18, 2024}{Washington, DC, USA}
\acmBooktitle{Proceedings of the 47th International ACM SIGIR Conference on
Research and Development in Information Retrieval (SIGIR '24), July 14--18,
2024, Washington, DC, USA}
\acmDOI{10.1145/3626772.3657954}
\acmISBN{979-8-4007-0431-4/24/07}

\begin{document}
\title{R-ODE: Ricci Curvature Tells When You Will be Informed}




\author{Li Sun}
\affiliation{%
  \institution{North China Electric Power University}
  \city{Beijing}
  \country{China}}
\email{ccesunli@ncepu.edu.cn}

\author{Jingbin Hu}
\affiliation{%
  \institution{North China Electric Power University}
  \city{Beijing}
  \country{China}}
\email{jingbinhu@ncepu.edu.cn}

\author{Mengjie Li}
\affiliation{%
  \institution{North China Electric Power University}
  \city{Beijing}
  \country{China}}
\email{mengjie@ncepu.edu.cn}

\author{Hao Peng}
\affiliation{%
  \institution{Beihang University}
  \city{Beijing}
  \country{China}}
\email{penghao@buaa.edu.cn}










\begin{abstract}
Information diffusion prediction is fundamental to understand the structure and organization of the online social networks, and plays a crucial role to blocking rumor spread, influence maximization, political propaganda, etc.  
So far, most existing solutions primarily predict the next user who will be informed with historical cascades, 
but  ignore an important factor in the diffusion process --- \textbf{the time}. 
Such limitation motivates us to pose the problem of the time-aware personalized information diffusion prediction for the first time, telling the time when the target user will be informed. 
In this paper, we address this problem from a fresh geometric perspective of Ricci curvature, and propose a novel Ricci-curvature regulated Ordinary Differential Equation (R-ODE).
In the diffusion process,  R-ODE  considers that the inter-correlated users are organized in a dynamic system in the representation space, and the cascades give the observations sampled from the continuous realm.
At each infection time, the message diffuses along the largest Ricci curvature, signifying less transportation effort.
In the continuous realm, the message triggers users' movement, whose trajectory  in the space   is parameterized by an ODE with graph neural network.
Consequently, R-ODE predicts the infection time of a target user by the movement trajectory learnt from the observations.
Extensive experiments evaluate the personalized time prediction ability of R-ODE, and show R-ODE outperforms the state-of-the-art baselines.
\end{abstract}

\ccsdesc[500]{Information systems~Social networks}
\ccsdesc[300]{Computing methodologies~Neural networks}

\keywords{Information Diffusion Prediction, Personalized Infection Time, Neural ODE, Ricci Curvature, Graph Neural Networks}




\maketitle

\section{Introduction}
Nowadays, online social networks have become indispensable and emerged as the predominant platform where the users share and exchange information with each other.
Information diffusion prediction primarily aims at identifying the next user who will be informed with the previous observations, i.e., historical cascades, and is receiving continuous attention \cite{Li2016DeepCasAE,Cao2017DeepHawkesBT}.
It has a series of applications ranging from influence maximization \cite{DeepInf}, public opinion analysis \cite{Rumor_1,Rumor_2,Rumor_3} to online advertising and recommendation \cite{Recommendation}. 

In the literature, early solutions on information diffusion prediction explore the sequential pattern in the historical cascades to predict the next infected user.
Recursive architectures such as LSTM and GRU are frequently revisited \cite{FOREST,deepdiffuse,DisenIDP}.
With the advancement of graph neural networks (GNNs) \cite{Kipf2016SemiSupervisedCW,Velickovic2017GraphAN}, the graph structure of social network has been incorporated  with historical cascades for better prediction \cite{DyHGCN,MS-HGAT}, achieving encouraging results in recent years.
Nevertheless, the existing solutions so far focus on predicting the probability of the user getting informed, and neglect an important factor in the diffusion process --- \textbf{the time}.
In fact, the time information is crucial to downstream applications, e.g., it is much easier to strengthen the propaganda or cut off the misinformation cascade if the infection time of the target individuals can be estimated.
To bridge this gap, we rethink information diffusion in a finer granularity, and propose the \textbf{time-aware personalized information diffusion prediction, supporting to predict the time when a target user will be informed.}
Note that, Deepdiffuse \cite{deepdiffuse} and the recent FOREST \cite{FOREST} allow for predicting the time when the next user will be informed, casting as a special case of the proposed problem.
To the best of our knowledge, we make the first attempt to predict the infection time of any possible user.

In this work, we approach information diffusion prediction from a fresh geometric perspective of graph Ricci curvature.
In Riemannian geometry, \textbf{Ricci curvature} quantifies the transportation effort from one node to another over the graph \cite{Ollivier2010ASO}, which is well aligned with the information diffusion from one to another.
Our intuition is that a message tends to infect the user of less transportation effort. 
Accordingly, we propose a novel Ricci-curvature regulated Ordinary Differential Equitation (\textbf{R-ODE}) for time-aware personalized prediction.
 Our idea is that, in the diffusion process, social network users as a whole is modeled as a dynamic system  in the representation space where \emph{the message diffuses along the largest Ricci curvature at each infection, and triggers the users' movement in the continuous realm.}
 Specifically, we propose a recursive graph convolution to obtain user/message coordinates (embeddings) at the time of each infection. 
 In the continuous realm, we model the trajectory of users' movement via a neural ODE built with GNN.
We interplay the system snapshots at each time points and the continuous realm to learn the movement trajectory, so that we are capable of predicting the infection time of a target user.

\noindent \textbf{Contribution Highlights.} The key contributions are three-fold. 
\textbf{(1) Problem.} We formulate the problem of time-aware personalized information diffusion prediction, supporting the infection time prediction of a target user.
\textbf{(2) Methodology.} We propose a novel R-ODE for the first time introducing Ricci curvature to model information diffusion, to the best of our knowledge.
\textbf{(3) Experiment.} We evaluate the superiority of R-ODE with SOTA baselines, and examine the proposed components by ablation study.


\section{Related Work}

\textbf{Information Diffusion Prediction.} 
Early practices, such as Deepdiffuse \cite{deepdiffuse},  learns the sequential pattern from the historical cascades for the next infected user prediction.
Later, researchers incorporate social structure and the cascades for better performance.
DyHGCN \cite{DyHGCN} constructs a heterogeneous graph containing the underlying social relations.
NDM \cite{NDM} learns dynamic social graphs by extending the LSTM model.
MS-HGAT \cite{MS-HGAT} pays extra attention to the topic of the message and estimates the potential interest of the user.
DisenIDP \cite{DisenIDP} predicts next infected user on the hypergraphs constructed from the structural and cascade information.
Most of the existing methods utilize traditional similarity measure in the prediction, and lack the ability to predict the infection time, which is also significant to downstream tasks.
Very recently, FOREST \cite{FOREST} conducts both next user and infection time prediction.
However, FOREST focus on predicting the time of the next infection occurrence,  and is still inadequate for the real case.
A more practical problem is to predict the time when the target individuals will be informed, but it has rarely been touched yet in the literature. 

\noindent \textbf{Neural Ordinary Differential Equations (ODEs).}
Neural ODEs \cite{nips18NODE} model the continuous dynamics by combining the principles of ODEs and neural networks.
Recent studies also make effort to extend neural ODE on the graphs.  
\citet{kdd20ND} combined ODEs and GNNs to learn the continuous-time dynamics of a graph. 
LG-ODE \cite{LG-ODE} and ODE-RNNs \cite{ODE-RNNs} are suitable for irregular time series. 
\citet{DBLP:conf/kdd/0002S023} introduce a generalized graph ODE to model the dynamics across environments.
\citet{HVGNN,aaai22SelfMix,SunL23AAAI,SunL24AAAI} demonstrate the expressiveness of generic manifold, especially for modeling graphs.
Recently, \citet{arXiv23chen,nips20lou,SunL24WWW} generalize ODEs on the manifold.
However, few researchers are devoted to design a specialized ODE for information diffusion. 

\begin{figure}
\centering
    \includegraphics[width=0.45\textwidth, height=0.22\textwidth]{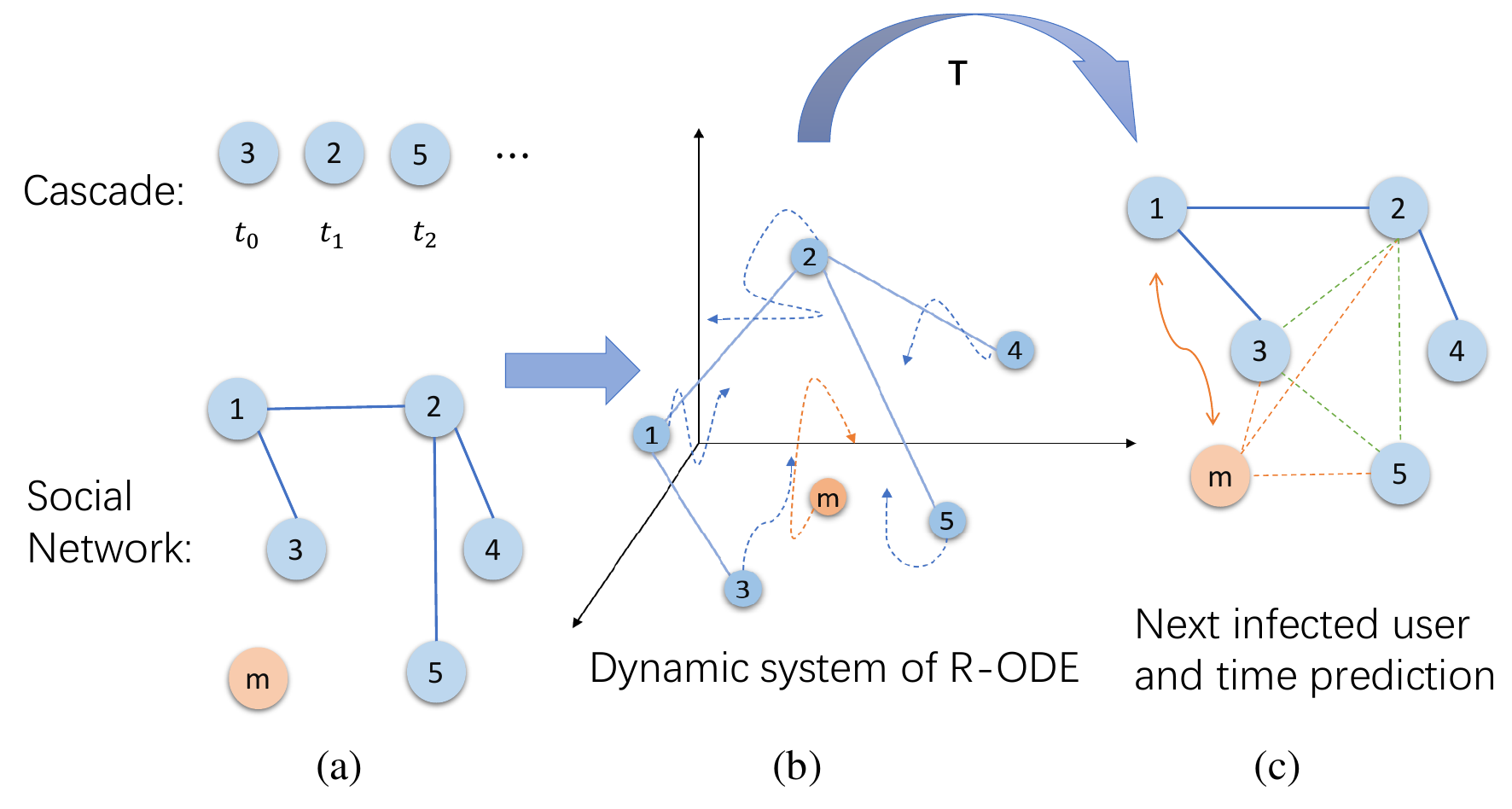}
        \vspace{-0.1in}
     \caption{Overall Architecture of R-ODE.}
    \label{main}
            \vspace{-0.15in}
\end{figure}

\section{Methodology}


\subsection{Preliminaries \& Problem Statement}

A social network is described as $G(\mathcal V, \mathcal E, \mathbf A, \mathbf X)$ over the user set $\mathcal V=\{u\}$ and the user linkage set $\mathcal E \subset \mathcal V \times \mathcal V$.
The number of users is $N$, and $\mathbf A$ is the N-by-N adjacency matrix.
Each user $u_i$ is attached a feature vector $\boldsymbol x_i$, recorded in $\mathbf X$.
For the message $m$, its cascade $C_m$ of length $L_m$ is a sequence of tuples $(u^k, t^k)$, denoting that user $u^k$ is affected at the time point $t^k$. 
 $C_m$ is collected in $\mathcal C$.

Distinguishing from majority studies in the literature, 
we highlight the time information in the diffusion process, and allow for predicting \emph{the time when the target individuals will be informed}.
Formally, we propose the problem of \textbf{time-aware personalize information diffusion prediction}.
Given a $G(\mathcal V, \mathcal E, \mathbf A, \mathbf X)$ and 
the historical cascades collected in $\mathcal C$,
we aim to (1) predict the users will be informed in the future, and (2) learn a mapping $\mathbf \Phi: (u, m) \to t$ for some message $m$, so that $t$ is the infection time point when the given user $u \in \mathcal V$ will be infected by the message $m$.
In this paper, we propose to study message-user infection from a new perspective of \textbf{Riemannian geometry} (i.e., the Ricci curvature on the graph), rather than traditional similarity measures of previous solutions.

\subsection{Overall Architecture}

To address this problem, we present a Ricci-curvature regulated ODE (\textbf{R-ODE}).
In brief, we model the social network as \textbf{a dynamic system in the representation space}, where the user's behaviors are inter-correlated to each other.
In \textbf{Sec 3.3}, we obtain the system snapshots (user/message coordinates in the space) at each infection times from structural information and historical cascades.
In \textbf{Sec 3.4}, the message tends to diffuse over the graph according to Ricci curvature, and trigger the movement of user nodes in the continuous realm. 
In \textbf{Sec 3.5}, we learn the continuous movement trajectory with the observations of system snapshots, so that we correctly predict the infection time of the target individuals.
The novelty is in two-fold. 
(1) We introduce the graph Ricci curvature to study message-user infection in the diffusion process.
(2) We design a neural ODE with GNN to parameterize the movement trajectory in continuous realm.
The overall architecture is illustrated in Figure \ref{main}. 

\subsection{Recursive Graph Convolution}

We propose a recursive graph convolution to obtain user coordinates (i.e., user embeddings $\boldsymbol h$) and message coordinate (i.e., message embedding $\boldsymbol m$) in the representation space at each infection time.
\emph{First, we introduce a temporal user-message graph} $G_m$ for a historical cascade $C_m$.
Concretely, $G_m$ is a weighted graph, whose node set is the union of user nodes $\mathcal V$ and the message node $m$, and the temporal graph structure is successively induced from the tuples in the cascade.
 For each tuple $(u^k, t^k)$, we add message-user and user-user links to the previous structure at time $t^{(k-1)}$. The message-user link connects the message node to user node $u^k$, while user-user link connects $u^k$ to the previously infected users in the cascade.
 
\emph{Second, we design a temporal attention mechanism} on $G_m$ where the pairwise attentional weight $w$ is derived from historical embeddings and the time information,
\vspace{-0.03in}
\begin{equation}
w_{ij}=\delta(\operatorname{MLP}((\boldsymbol h_i^{k-1}+\boldsymbol h_j^{k-1})||(\boldsymbol t^{k}+\boldsymbol t^{k-1}))),
\vspace{-0.05in}
\label{weight}
\end{equation}
where we slightly abuse the notation $\boldsymbol h_j^{k-1}$ for both user and message embedding at the prior time point.
$\delta$, $\operatorname{MLP}$ and $||$ denote the sigmoid activation, multi-layer perception and vector concatenation, respectively.
In Eq. \ref{weight}, we utilize time encoding $\boldsymbol t^k$ rather than the scalar time points. 
In practice, we adopt the translation-invariant encoding \cite{SunL22CIKM}, which encodes the time point $t^k$ as a vector encoding $\boldsymbol t^{k} \in \mathbb R^T$ with the following map
\vspace{-0.03in}
\begin{equation} \label{time.encoder}
        \boldsymbol t^{k} = \sqrt{\frac{1}{T}}\left[\cos\left(\omega_1 t^k + \theta_1\right), \cdots , \cos\left(\omega_T t^k + \theta_T\right)\right], 
        \vspace{-0.05in}
\end{equation}
where $\omega$'s and $\theta$'s are the learnable parameters to construct the time encoding.
\emph{Third, we recursively update user embedding with attentive aggregation every time a new user is infected}, and we have
    \vspace{-0.03in}
\begin{equation}
    {\boldsymbol h}_{i}^{k}=\delta({\boldsymbol h}_{i}^{k-1}+\sum\nolimits_{j \in \mathcal N_i} {w}_{ij}{\boldsymbol h}_{j}^{k-1}), 
    \vspace{-0.05in}
\end{equation}
where $j \in \mathcal N_i$ denote the neighboring user of user $i$ on $G_m$. 
Meanwhile, we successively refine the message's coordinate by computing the weighted centroid of user embeddings in the space,
\begin{equation}
    {\boldsymbol m}^{k}=\delta ({\boldsymbol m}^{k-1}+\sum {w}_{mi}{\boldsymbol h}_{i}^{k}),
\end{equation}
where we have $\boldsymbol m^0=\boldsymbol h^0$ in particular. 
In our design, we apply a graph convolution layer \cite{Kipf2016SemiSupervisedCW} on $G$ to derive user embedding $\boldsymbol h^0$ in the representation space, before studying the coordinate evolvement in the message diffusion process.

\subsection{Diffusion along Graph Ricci Curvature}

In Riemannian geometry, \textbf{Ricci curvature} \cite{Ollivier2010ASO,SunL23IJCAI} quantifies the transportation effort from one node to another node over the graph.
\emph{Intuitively, the message tends to infect the node with minimal transportation effort}, and thus diffuse along the direction of maximum Ricci curvature accordingly.
Given a weighted $G_m$ at $t^k$, the Ricci curvature between message and user is defined as follows,
\vspace{-0.05in}
\begin{equation}
    {Ric}(m, u_i; t^k)=1-\frac{W({p}^k_{m},{p}^k_{i})}{d(\boldsymbol m^k, \boldsymbol h^k_i)},
    \vspace{-0.05in}
\end{equation}
where ${p}^k_{m}$ and ${p}^k_{i}$ are the mass distribution surrounding the message node and user, respectively.
The mass distribution is given as 
\vspace{-0.05in}
\begin{equation}
    {p}^k_{i}(j)=\begin{cases}
\alpha ,& j=i\\ 
\frac{1}{degree_i}(1-\alpha),& j\in \bar{\mathcal{N}}_{i},
\end{cases}
\vspace{-0.05in}
\end{equation}
and $\bar{\mathcal{N}}_{i}$ is the neighbor node set of $i$ at time $t^k$, $\alpha$ is a hyperparameter.
$W$ is the Wasserstein distance between the mass distributions, while $d$ is the distance in the representation space.
The challenge is that \textbf{Wasserstein distance} is typically calculated by solving a linear programming, and thus \textbf{prevents the gradient backpropagation of the neural model} \cite{Ye2020CurvatureGN}.
To bridge this gap, we propose a differentiable surrogate with Kantorovich-Rubinstein duality \cite{Gulrajani2017ImprovedTO}, 
\vspace{-0.05in}
\begin{equation}
    \begin{aligned}
        W(p^k_m, p^k_i) 
        & = \sup \nolimits_{\lVert f \rVert_L \leq 1} \mathbb{E}_{z\sim p^k_m}[f(z)] - \mathbb{E}_{z\sim q^k_i}[f(z)] \\
        &= \sup \nolimits_{\lVert f \rVert_L \leq 1} \sum\nolimits_{j \in \bar{\mathcal{V}}} f(j)p^k_m(j) - \sum\nolimits_{j \in \bar{\mathcal{V}}} f(j)p^k_i(j) ,
    \end{aligned}
    \vspace{-0.02in}
    \label{sup}
\end{equation}
where $f$ is a $1-$Lipschitz continuous function. $ \bar{\mathcal{V}}$ is the node set of $G_m$.
Thus a \textbf{surrogate Wasserstein distance} is derived as follows,
\begin{equation}
   \hat{W}(p^k_m, p^k_i) =[\boldsymbol L^k f(\boldsymbol H^k)]_m-[\boldsymbol L^k f(\boldsymbol H^k)]_i ,
 \label{Ric}
  \vspace{-0.03in}
\end{equation}
where $f: \mathbb R^{(N+1)\times d} \to \mathbb R^{(N+1)}$ takes the form of affine transform, satisfying  $1-$Lipschitz condition \cite{SunL23ICDM}. 
At the time $t^k$, we have $\boldsymbol L=\alpha \mathbf I +(1-\alpha)(\mathbf D^k)^{-1}\mathbf A^k$, where $\mathbf A^k$ and $\mathbf D^k$ are the adjacency matrix and diagonal degree matrix, respectively.
User and message embeddings  are recorded in $\boldsymbol H^k$, and $[\cdot]_i$ denotes the $i$-th element of the vector.
As a result, the probability of a user getting infected $p(u_{\operatorname{inf}}=u_j;t^k)$ is given by a softmax function among Ricci curvatures $Ric(m,u_i), u_i \in \mathcal V$, and a loss of likelihood maximization is
\vspace{-0.1in}
\begin{equation}
   \mathcal L_{Ricci}=- \frac{1}{\sum_{C_m \in \mathcal C} L_m}\sum\nolimits_{C_m \in \mathcal C} \sum\nolimits_{k=1}^{L_m}\log p(u_{\operatorname{inf}}=u^k;t^k).
 \label{Ric}
 \vspace{-0.03in}
\end{equation}
\subsection{Modeling Dynamics with Graph ODE}

Here, we build a neural ODE with GNN to model the underlying dynamics in the continuous realm, supporting the infection time prediction.
In the dynamic system, the injected message triggers the users' movement as shown in Figure \ref{main} (c), and we propose to study the movement by the velocity vector. 
We briefly review some important notions. The velocity vector $v$ is derived by differentiating movement trajectory $\phi$ regarding time and, accordingly, the (velocity) vector field of the system is described by the \textbf{ODE},
\vspace{-0.05in}
\begin{equation}
\frac{\partial }{\partial t_{\operatorname{sys}}}\phi(u_i, t_{\operatorname{sys}})=v(\phi(u_i, t_{\operatorname{sys}}),t_{\operatorname{sys}}), \quad \phi(u_i, 0)=\boldsymbol h^1_i,
 \vspace{-0.05in}
\end{equation}
where $\phi(u_i, 0)$ denotes the user's initial position in the system, and the time $t_{\operatorname{sys}}$ is scaled to $[0,1]$  without loss of generality.
In R-ODE, \textbf{we parameterize  the vector field $v$ as a neural net, and the vector net is built with a GNN and time encoding} as follows,
\vspace{-0.05in}
\begin{equation}
v(\phi(u_i, t_{\operatorname{sys}}),t_{\operatorname{sys}})= \operatorname{MLP}(g(u_i) || {\boldsymbol t}_{\operatorname{sys}}), \quad g(u_i)=\operatorname{GNN}(\mathbf A, \mathbf X),
\label{gnn}
 \vspace{-0.03in}
\end{equation}
where the time encoding ${\boldsymbol t}_{\operatorname{sys}}$ is given by Eq. (\ref{time.encoder}). 
The GNN-based vector net is a key ingredient of our design, 
and the rationale is that user's movement is not only determined by its own feature but also affected by the neighboring nodes.
To model the inter-correlation among the users, we propose to nest a GNN on $G(\mathcal V, \mathcal E, \mathbf A, \mathbf X)$ in the velocity net as in Eq. (\ref{gnn}), and the GNN can be instantiated with any popular architecture, e.g., GCN.
(We will show the superiority of GNN-based vector net  in Ablation Study).
The movement trajectory is thus given by the well-established ODE solvers \cite{Dormand2017NumericalMF},  $\phi(u_i, t_{\operatorname{sys}})=\texttt{ODE\_Solver}(\phi(u_i, 0), [0,  t_{\operatorname{sys}}], v)$.
We learn the trajectory with the observations at each infection time, and obtain the loss of mean squared error as follows,
\vspace{-0.05in}
\begin{equation}
\mathcal L_{ODE}=\frac{1}{\sum_{C_m \in \mathcal C} L_m}\sum\nolimits_{C_m \in \mathcal C} \sum\nolimits_{k=1}^{L_m}||\phi(u_i, t^k_{\operatorname{sys}}) - \boldsymbol h_i^k ||^2,
 \vspace{-0.05in}
\end{equation}
and $t^k_{\operatorname{sys}}$ is the rescaled time by the maximum value in each cascade.
\textbf{R-ODE is learnt by the overall objective of }$\mathcal L_{Ricci}+\mathcal L_{ODE}$.  R-ODE tells the time when the target user encounters the message in the representation space with the movement trajectory.

\begin{table*}
  \vspace{-0.1in}
\caption{Next infected user prediction results on Twitter, Android and  Christianity datasets in terms of H@K and M@K (\%). }

  \vspace{-0.1in}
\begin{tabular}{lcccccccccccc}
                                 & \multicolumn{4}{c}{Twitter}                        & \multicolumn{4}{c}{Android}                      & \multicolumn{4}{c}{Christianity} \\
\multicolumn{1}{l|}{}            & H@10  & M@10  & H@50  & \multicolumn{1}{c|}{M@50}  & H@10  & M@10 & H@50  & \multicolumn{1}{c|}{M@50} & H@10   & M@10   & H@50   & M@50  \\ \hline   \hline
\multicolumn{1}{l|}{Deepdiffuse} & 5.79  & 5.87  & 10.80 & \multicolumn{1}{c|}{6.80}  & 4.13  & 2.30 & 10.58 & \multicolumn{1}{c|}{2.53} & 10.27  & 7.27   & 21.83  & 7.83  \\
\multicolumn{1}{l|}{FOREST}      & 25.15 & 16.81 & 38.27 & \multicolumn{1}{c|}{17.40} & 8.71  & 5.81 & 16.40 & \multicolumn{1}{c|}{6.13} & 22.88  & 14.63  & 41.22  & 15.41 \\
\multicolumn{1}{l|}{Inf-VAE}     & 14.73 & 19.79 & 32.72 & \multicolumn{1}{c|}{18.66} & 5.55  & 4.82 & 12.96 & \multicolumn{1}{c|}{4.70} & 17.20  & 8.39   & 36.11  & 10.73 \\
\multicolumn{1}{l|}{DyHGCN}      & 27.91 & \underline{20.68} & 46.42 & \multicolumn{1}{c|}{17.52} & 6.69  & 3.51 & 16.06 & \multicolumn{1}{c|}{3.91} & 24.10  & 12.06  & 42.64  & 13.09 \\
\multicolumn{1}{l|}{NDM}         & 22.21 & 12.43 & 28.23 & \multicolumn{1}{c|}{13.21} & 4.83  & 2.01 & 14.24 & \multicolumn{1}{c|}{2.23} & 15.41  & 7.42   & 31.36  & 7.68  \\
\multicolumn{1}{l|}{MS-HGAT}     & \underline{30.93} & 18.67 & \underline{46.67} & \multicolumn{1}{c|}{\underline{19.38}} & \underline{10.33} & \underline{6.35} & \underline{21.20} & \multicolumn{1}{c|}{\underline{6.78}} & 28.99  & 17.08  & 46.94  & 17.89 \\
\multicolumn{1}{l|}{DisenIDP}    & 27.21 & 20.22 & 46.22 & \multicolumn{1}{c|}{14.10} & 9.55  & 6.32 & 19.34 & \multicolumn{1}{c|}{6.70} & \underline{31.20}  & \underline{18.99}  & \underline{50.66}  & \underline{19.91} \\ 
\hline \hline
\multicolumn{1}{l|}{\textbf{R-ODE} (Ours) }      & \textbf{31.65}     & \textbf{21.35}     & \textbf{48.46}     &  
 \multicolumn{1}{c|}{\textbf{22.31}}     & \textbf{10.89}     & \textbf{6.74}   & \textbf{23.25}     & \multicolumn{1}{c|}{\textbf{7.21}}    & \textbf{31.84}      & \textbf{20.64}      & \textbf{51.46}      & \textbf{20.99}    
\end{tabular}
\label{results}
  \vspace{-0.1in}
\end{table*}

\section{Experiment}
\subsection{Experimental Setups}
\noindent\textbf{Datasets \& Baselines.}
We evaluate the proposed R-ODE on three benchmark datasets, i.e., Twitter \cite{twitter}, Android \cite{Inf-VAE} and Christianity \cite{Inf-VAE}.
Both time-unaware baselines (Inf-VAE \cite{Inf-VAE},
DyHGCN \cite{DyHGCN},
NDM \cite{NDM},
MS-HGAT \cite{MS-HGAT}
and
DisenIDP \cite{DisenIDP})
and time-aware baselines (Deepdiffuse \cite{deepdiffuse} and the recent 
FOREST \cite{FOREST}) are compared.
None of the existing solution consider time-aware personalized diffusion prediction on the graphs, to the best of our knowledge, and the proposed R-ODE bridges this gap.

\noindent\textbf{Evaluation Metrics.}
 In the diffusion process, we consider the tasks of (1) the next infected user prediction and (2) personalized infection time prediction.
For the former, we follow the settings of previous studies \cite{FOREST,MS-HGAT,Inf-VAE}, and employ the evaluation metrics of mean average precision on top K (\textbf{M@K}) and Hits scores on top K (\textbf{H@K}).
For the latter, we utilize the metric of Root Mean Square Error (\textbf{RMSE}) same as \citet{deepdiffuse}.
We report the mean value of $5$ independent runs for both learning tasks.


\noindent\textbf{Reproducibility \& Model Configuration.}
In  R-ODE, the affine transform of the surrogate Wasserstein distance is instantiated as a linear layer without activation. 
In Ricci loss, we suggest to mask $Ric(m, u_i)$ of the prior infected user in the cascade before calculating the infection probability at the current time.
In the velocity net, we stack GCN layer \cite{Kipf2016SemiSupervisedCW} once, and utilize a MLP with $2$ hidden layers.
The dropout is set as $0.3$. The dimension of time encoding is $16$.

\begin{figure}[t]
  \vspace{-0.05in}
\centering 
\subfigure[Infection Time]{
\includegraphics[width=0.48\linewidth]{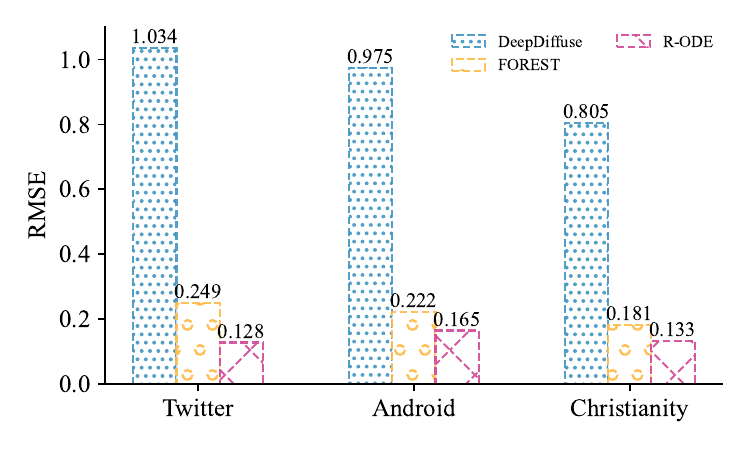}}
  \hspace{-0.05in}
\subfigure[Next Infected User]{
\includegraphics[width=0.48\linewidth]{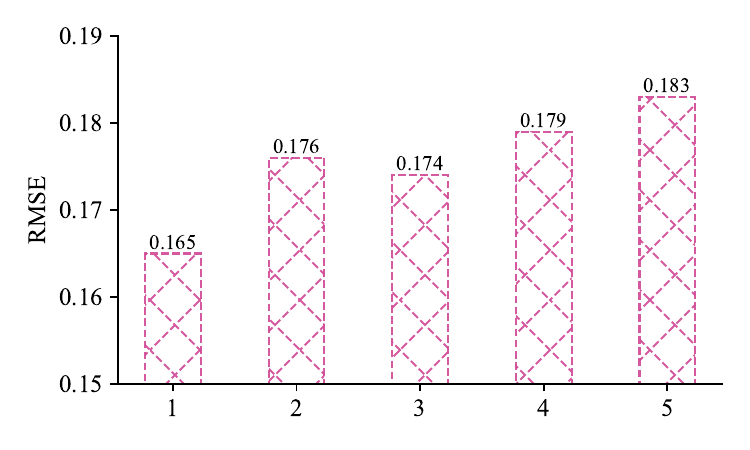}}
  \vspace{-0.1in}
\caption{Results of infection time prediction on Android.}
\label{fig2}
  \vspace{-0.2in}
\end{figure}
  
\subsection{Results and Analysis}

\noindent\textbf{Next Infected User Prediction. }
The results in terms of M@K and H@K is collected in Table \ref{results}, where K=$\{10, 50\}$. 
The existing methods typically returns a candidate list of the next infected user, order by the similarity measure.
Instead the traditional measures,  R-ODE selects the users getting infected by value of message-to-user Ricci curvature.
As shown in Table \ref{results}, the proposed R-ODE consistently achieves the best result on all the datasets, e.g.,  H@10 of R-ODE is more than $5 \times$ to that of Deepdiffuse. A reason is that the Ricci curvature better models the diffusion direction, 
and we will further discuss the Ricci curvature in Ablation Study.

\begin{table}
\caption{Ablation study on Android dataset. }
\label{ablation}
\vspace{-0.12in}
\begin{tabular}{l | ll|ll|l}
      & H@10 & M@10 & H@50 & M@50 & RMSE \\ \hline  \hline
\textbf{R-ODE} &\textbf{10.89}      &\textbf{6.74}      &\textbf{23.25}      &\textbf{7.21}      &\textbf{0.165}  \\
$-GNN$ &6.77      &5.08      &18.31      &5.03      &0.869      \\
$-Ricci$  &6.98      &4.21      &16.55      &4.59      &0.716      \\
$-R/GNN$  &5.51      &3.53      &12.67      &3.64      &0.915      \\   
\end{tabular}
\vspace{-0.1in}
\end{table}

\noindent\textbf{Infection Time Prediction.}
We compare with the few time-aware baselines, i.e., Deepdiffuse and FOREST, and show the RMSE results on Android dataset in Figure \ref{fig2} (a).
Our R-ODE achieves the minimal error regarding the groundtruth, and it suggests that the designed ODE captures the dynamics of the diffusion process.
Moreover, we evaluate the infection time of the future $k-$th user in the cascade, and report the RMSE in Figure \ref{fig2} (b).
Note that, neither Deepdiffuse or FOREST can tell the infection time of a given future user in the cascade.
It shows that R-ODE is able to effectively predict the infection time of the 3rd or 4th infected user in the future.


\noindent\textbf{Ablation Study.}
Here, we examine the two key ingredients of the proposed R-ODE: (1) Ricci curvature-based diffusion, and (2) GNN-based velocity net in ODE.
The $-Ricci$ variant replaces the message-to-user Ricci curvature by cosine similarity in calculating the infection probability.
The $-GNN$ variant replaces the one layer GNN by a linear layer with activation, neglecting the inter-correlation among the users.
The $-R/GNN$ variant replaces both  Ricci curvature-based diffusion and GNN of R-ODE.
The results of next infected user prediction and infection time prediction are reported in Table \ref{ablation}.
We find that $-GNN$ variant achieves inferior performance to R-ODE.
It shows that Ricci curvature, quantifying the transportation effort from one node to another, tends to be more effective than the traditional cosine similarity, verifying the motivation of our work.

\section{Conclusion}

Herein, we propose the problem of the time-aware personalized information diffusion prediction, and present a neural model of  R-ODE, combining the advantages of neural ODE and Ricci curvature of Riemannian geometry.
In R-ODE, we consider a user-message dynamic system in the representation space, where we first propose the recursive graph convolution to learn user/message coordinates at each infection time.
The message diffuses along Ricci curvature, and users move along the trajectory parameterized by a GNN-based ODE. 
Thus, we predict the infection time of target user with the trajectory.
Empirical results show the superiority of R-ODE.

\begin{acks}
This work is supported in part by NSFC through grants 62202164 and 62322202. Corresponding authors: Li Sun and Hao Peng.
\end{acks}

\balance
\bibliographystyle{ACM-Reference-Format}
\bibliography{SIGIR24_1}

\end{document}